\documentclass[runningheads]{llncs}
\usepackage[T1]{fontenc}
\usepackage{booktabs} 
\usepackage{colortbl} 
\usepackage{xcolor}
\usepackage{amsmath}
\usepackage{marvosym}
\usepackage{multirow}
\usepackage{graphicx,verbatim}
\usepackage{amsmath}
\usepackage{amssymb}
\usepackage{tabularx} 
\begin{document}
\title{PRAD: Periapical Radiograph Analysis Dataset and Benchmark Model Development}
\author{Zhenhuan Zhou\inst{1, 2} \and
Yuchen Zhang\inst{3} \and
Ruihong Xu\inst{1} \and
Xuansen Zhao\inst{4} \and\\
Tao Li\inst{1, 5}\textsuperscript{(\Letter)}}
\authorrunning{Z. Zhou et al.}
\institute{College of Computer Science, Nankai University, Tianjin 300350, China \and Key Laboratory of Data and Intelligent System Security, Ministry of Education, China \and Department of stomatology, Tianjin Union Medical Center, The First Affiliated Hospital of Nankai University,  Tianjin, 300121, China \and
College of Mechanical Engineering, Taiyuan University of Technology, Shanxi 030000, China \and Haihe Lab of ITAI, Tianjin 300459, China \\
\email{litao@nankai.edu.cn}
}

\maketitle
\begin{abstract}
Deep learning (DL), a pivotal technology in artificial intelligence, has recently gained substantial traction in the domain of dental auxiliary diagnosis. However, its application has predominantly been confined to imaging modalities such as panoramic radiographs and Cone Beam Computed Tomography, with limited focus on auxiliary analysis specifically targeting Periapical Radiographs (PR). PR are the most extensively utilized imaging modality in endodontics and periodontics due to their capability to capture detailed local lesions at a low cost. Nevertheless, challenges such as resolution limitations and artifacts complicate the annotation and recognition of PR, leading to a scarcity of publicly available, large-scale, high-quality PR analysis datasets. This scarcity has somewhat impeded the advancement of DL applications in PR analysis. In this paper, we present PRAD-10K, a dataset for PR analysis. PRAD-10K comprises 10,000 clinical periapical radiograph images, with pixel-level annotations provided by professional dentists for nine distinct anatomical structures, lesions, and artificial restorations or medical devices, We also include classification labels for images with typical conditions or lesions. Furthermore, we introduce a DL network named PRNet to establish benchmarks for PR segmentation tasks. Experimental results demonstrate that PRNet surpasses previous state-of-the-art medical image segmentation models on the PRAD-10K dataset. The codes and dataset will be made publicly available.


\end{abstract}

\section{Introduction}
\begin{figure}[t]
\includegraphics[width=\textwidth]{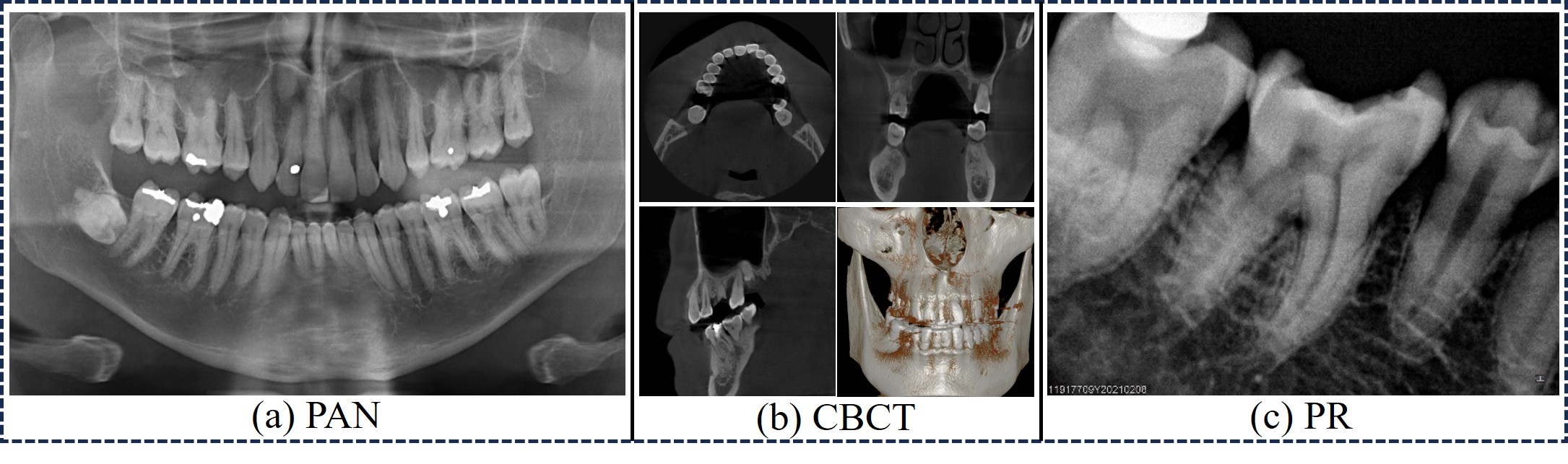}
\caption{Examples of three dental radiographic imaging modalities. (a) PAN, (b) CBCT and (c) PR.} \label{fig1}
\end{figure}

Radiological evaluation plays a pivotal role in dentistry, enabling dentists to diagnose conditions, formulate treatment plans, and assess treatment outcomes through various imaging techniques \cite{lo2018accuracy}. As illustrated in Fig. \ref{fig1}, the primary dental radiographic imaging modalities are categorized into three types: Panoramic Radiography (PAN), Cone Beam Computed Tomography (CBCT), and Periapical Radiography (PR). Among these, PR is the most frequently employed imaging technique in endodontics, particularly for diagnosing apical periodontitis and conducting root canal treatments. PR is an intraoral imaging method where the sensor is strategically positioned within the patient's mouth to capture images, facilitating a detailed examination of one or two teeth, including pulp, apical, and periodontal conditions \cite{stera2024diagnostic,patel2009new}. While PAN can provide a comprehensive view of the dental arch, it lacks the localized detail that PR offers \cite{tiburcio2021global}. Conversely, CBCT, as a 3D imaging modality, delivers detailed visualization of teeth and surrounding bone structures. However, its high cost and increased radiation exposure make it less preferable in endodontics \cite{shokri2024cone,tiburcio2021global}. Therefore, leveraging Deep Learning (DL) techniques for the intelligent analysis of PR can significantly enhance Computer-Aided Diagnosis (CAD) in endodontics.

Recently, there has been a surge of research and applications of DL in the field of dentistry. For instance, Wang et al. \cite{wang2023root} introduced a deep multi-task learning framework designed to segment teeth and root canals in dental CBCT images. Jang et al. \cite{jang2024fully} developed a fully automated system for integrating intraoral scans (IOS) with CBCT through individual tooth segmentation and identification. Zhang et al. \cite{zhang2023children} created a pediatric PAN dataset aimed at tooth segmentation and disease detection. Cui et al. \cite{cui2022fully} proposed a fully automated method that achieves precise tooth and bone instance segmentation in CBCT images by integrating dual features of tooth centroids and skeletons. The WS-TIS network \cite{wang2024weakly} employs weakly supervised tooth instance segmentation with multi-label learning, facilitating accurate segmentation of teeth in dental 3D models. Additionally, Chen et al. \cite{chen2024convolutional} and Ozsari et al. \cite{ozsari2025automatic} utilized DL for the early diagnosis of periodontal bone loss and the identification of vertical root fractures, respectively, using a small dataset of approximately 400 PR images. These studies suggest that, although the application of DL in assisting dental diagnosis is increasingly growing and holds great potential, its application in PR diagnosis within endodontics remains limited.

In this paper, we introduce PRAD-10K, a large-scale PR dataset featuring expert annotations, which serves as a potential benchmark for research in DL-based PR image analysis. Detailed information about the dataset is provided in Section \ref{sec2}. Additionally, to tackle the multi-scale challenges inherent in PR image segmentation tasks, we present PRNet, a DL network that integrates the Multi-scale Wavelet Convolution Network (MWCN) and the Channel Fusion Attention (CFA) mechanism. Extensive experiments reveal that PRNet surpasses previous state-of-the-art (SOTA) medical image segmentation networks on the PRAD-10K dataset. 

In summary, the contributions of this paper are as follows: (1) We introduce PRAD-10K, a large-scale dataset for PR image analysis. (2) We propose PRNet, a DL segmentation network based on the MWCN and CFA mechanism, which effectively addresses the multi-scale challenges in PR segmentation tasks. (3) We conduct extensive experiments demonstrating that PRNet outperforms previous SOTA networks on the PRAD-10K dataset. Ablation experiments also demonstrated the effectiveness of each component of PRNet.

\section{PRAD-10K}\label{sec2}
\subsection{Overview}
\begin{figure}[t]
\includegraphics[width=\textwidth]{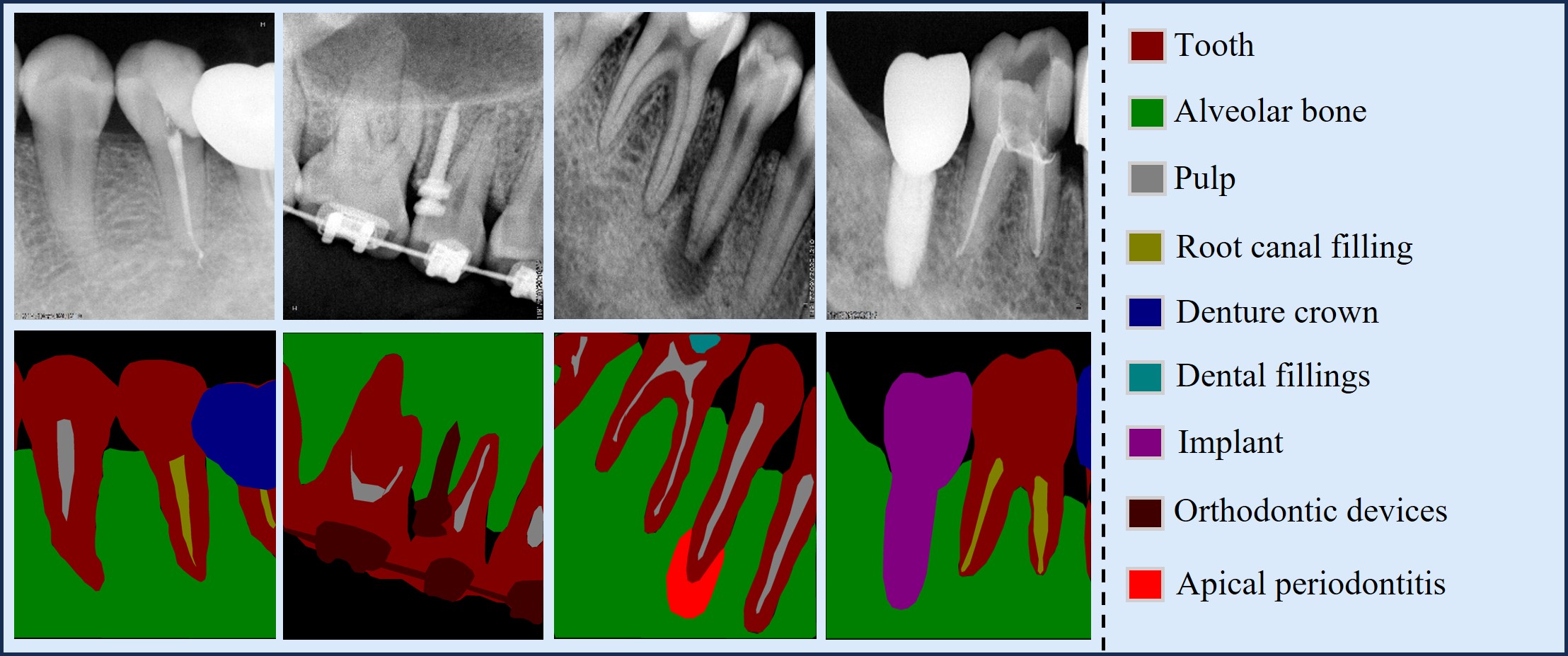}
\caption{Examples from the PRAD-10K dataset (top left) along with their corresponding pixel-level annotations (bottom left). The right side displays the label distribution.} \label{fig2}
\end{figure}

In Table \ref{tab1}, PRAD-10K is compared with some existing 2D dental datasets. Most prior datasets are small and focus on PAN or occlusal and intraoral radiographs, highlighting a gap in large-scale PR datasets. The introduction of PRAD-10K provides a benchmark for the application of DL in the analysis of PR images. Fig. \ref{fig2} shows four PRAD-10K images with expert pixel-level annotations, covering all category labels. On the right side of Fig. \ref{fig2}, a detailed label index is provided. PRAD-10K includes nine types of pixel-level annotations for anatomical structures, lesions, and restorations or devices. Additionally, classification labels are provided for some images with periodontitis, apical periodontitis, or inadequate root canal fillings.
\subsection{Collection and Annotation}

\begin{table}[t]
\caption{Comparison between PRAD-10K and some existing publicly available 2D dental datasets, OR and IR stands for Occlusal and Intraoral Radiograph, respectively.}
\label{tab1}
\centering
\begin{tabularx}{\textwidth}{>{\centering\arraybackslash}m{2cm} >{\centering\arraybackslash}m{2cm} >{\centering\arraybackslash}m{1.5cm} >{\centering\arraybackslash}m{1.5cm} >{\centering\arraybackslash}X}
\hline
Dataset & Modality & Year & Size & Task\\
\hline
PDSM \cite{abdi2015automatic} &PAN & 2020 & 116 & Mandibles segmentation\\

TDD \cite{panetta2021tufts} & PAN & 2021 & 1000 & Tooth segmentation\\

PRD \cite{roman2021panoramic} & PAN & 2021 & 598 & Tooth segmentation\\

OCI\cite{kaggleOC} & RGB Photo & 2022 & 131 & Oral cancer classification\\

DC1000\cite{wang2023multi} & PAN \& OR & 2023 & 597+2389 & Caries segmentation\\

IO150K\cite{zou2024teeth} & IR & 2024 & 1,500,000 & Tooth segmentation\\
\hline
PRAD-10K \vspace{-9pt} & PR\vspace{-9pt} & 2025\vspace{-9pt} & 10,000\vspace{-9pt} & Multi-structure segmentation Disease classification\\
\hline
\end{tabularx}
\end{table} 

PRAD-10K data are sourced from real clinical data from a top-tier hospital's endodontics department, ensuring no private information is included for research use. All research data in this study were approved by the ethics committee of the collaborating hospital, ensuring compliance with the Declaration of Helsinki. During the data collection phase, a radiology expert selected 10,000 high-quality PR images based on criteria such as clear teeth visibility, absence of severe artifacts, and exclusion of master cone or working length radiographs. Additionally, PR images with rich structures, such as dental implants, orthodontic treatments, or fillings, were intentionally selected to enhance the dataset's richness.

Two experienced endodontists and a computer researcher performed the data annotation work. The data were divided into two groups, each processed by a different endodontist. Subsequently, the computer researcher trained the two doctors to ensure they could correctly use the annotation software, Labelme \cite{russell2008labelme}. During the manual annotation process, the computer researcher regularly compiled the annotated data, checked the usability of the labels, and ensured the accuracy of the label index. After the initial annotation, the endodontists reviewed each other's work, discussing and correcting any issues. During this stage, the endodontists also assigned classification labels to images with confirmed lesions. Finally, the computer researcher reviewed all data for correct formatting. The dataset construction took over 8 months.

\section{Method}
\subsection{Overview}
\begin{figure}[t]
\includegraphics[width=\textwidth]{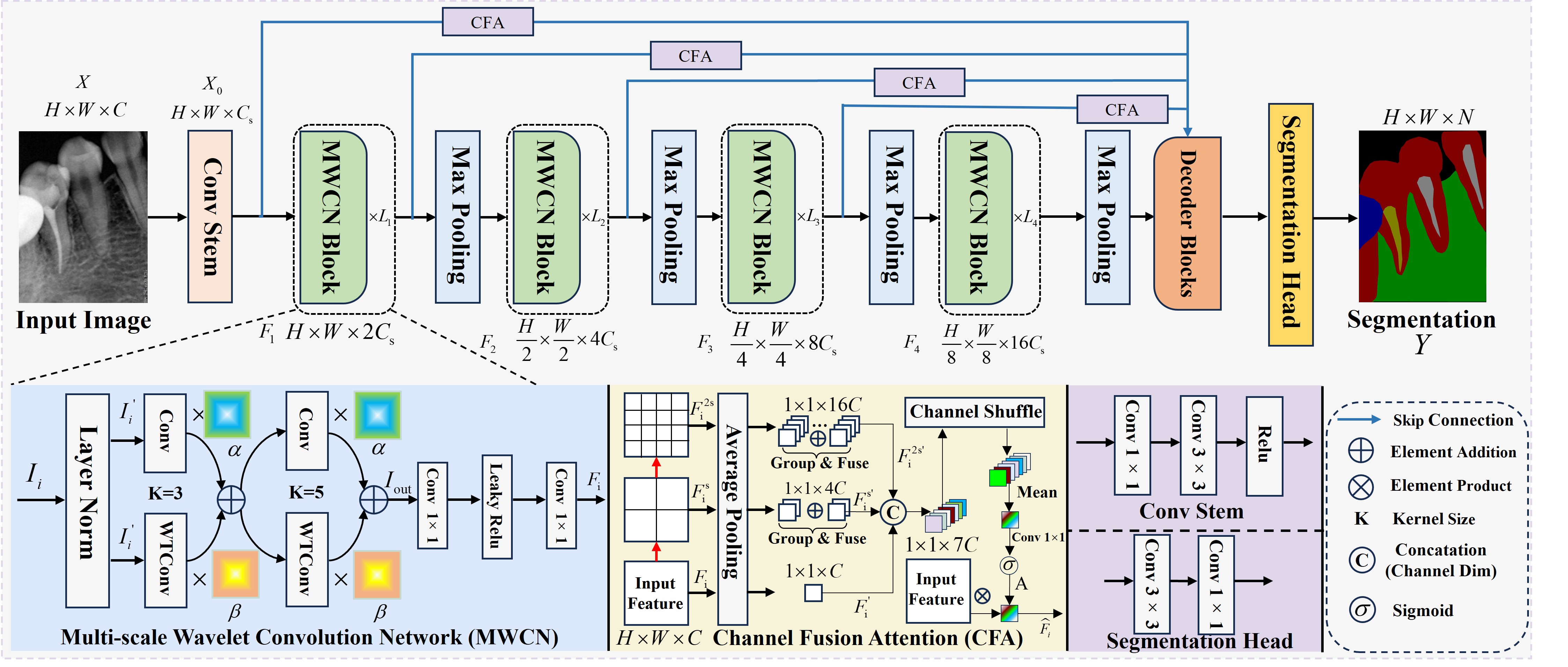}
\caption{Overall framework of the proposed PRNet. The structure of the decoder blocks are consistent with UNet.} 
\label{fig4}
\end{figure}

The overall pipeline of PRNet is illustrated in Fig. \ref{fig4}. It has two main components: MWCN blocks as the encoder and the CFA mechanism for skip connections. Specifically, given an input image $X \in \mathbb{R}^{H\times W \times C}$, here, $H$ and $W$ represent the height and width of the image, respectively, and $C$ represents the number of channels. The initial feature map $ X_0 \in \mathbb{R}^{H\times W \times C_s}$ for subsequent feature learning is obtained through a conv stem block. Next, $X_0$ is fed into the MWCN blocks, which consists of four stages, resulting in four hierarchical features $F_i \in \mathbb{R}^{\frac{H}{2^{i-1}} \times \frac{W}{2^{i-1}} \times 2i \cdot C_s}$, $i \in \{1, 2, 3, 4\}$. Each stage of the encoder is composed of $L_i$ MWCN blocks and a MaxPooling layer. Subsequently, $F_4$ is fed into the decoder blocks, while $X_0$, $F_1 $, $F_2$ and $F_3$ will first pass through the CFA mechanism to obtain weighted feature maps. These features are then fused with corresponding decoder features via skip connections. Finally, the decoder features are processed by a segmentation head to produce the segmentation mask $Y \in \mathbb{R}^{H\times W \times N}$, where N is the number of classes.

\subsection{Multi-scale Wavelet Convolution Network}
Inspired by \cite{finder2024wavelet}, WTConv's large receptive field is crucial for learning global knowledge in PR images. However, to ensure that PRNet captures both global features and local details (which is essential for addressing the multi-scale challenges in PR images), we designed the MWCN blocks. As shown in the blue section at the bottom left of Fig. \ref{fig4}, for a given input $I_{i} \in \mathbb{R}^{H\times W \times C}$,  MWCN includes two layers of feature extraction: two convolutional layers and two WTConv layers, both with kernel sizes of 3 and 5, respectively. The WTConv branches are responsible for capturing global features with a large receptive field, while the convolutional layers extract local details, complementing each other. For the outputs of each level's convolutional layers and WTConv layers, we use two independent Global-local Feature Weighting Matrices (GFWM) $\alpha, \beta \in \mathbb{R}^{H\times W \times 1}$ to weight and fuse the two outputs. This allows the network to adaptively and dynamically adjust the importance of global and local features during the fusion process. Finally, the fused features $I_{out}$ are then passed to the MaxPooling layer and the next stage of MWCNs through a simple feed-forward block. The overall process can be represented by Equation \ref{eq1}, where $I'_{i}$ represents the output of the input features $I$ after LayerNorm block.
\begin{equation}
\begin{aligned}
    I_{1}=\alpha \cdot(Conv_{k=3}(I'_{i}))+ \beta \cdot (WTConv_{k=3}(I'_{i}))\\ I_{2}=\alpha \cdot(Conv_{k=5}(I_{1}))+ \beta \cdot (WTConv_{k=5}(I_{1}))\\
    F_{i}=Conv_{k=1}(LeakyRelu(Conv_{k=1}(I_{2})))
\end{aligned}
\label{eq1}
\end{equation}

\subsection{Channel Fusion Attention}
The main structure of the CFA is shown in the yellow section at the bottom of Fig. \ref{fig4}. The primary function of the CFA is to weight the features from a channel perspective by integrating different levels of local features within the feature layers. This process enhances the decoder's ability to recognize objects of various sizes by feeding the feature layer, enriched with multi-scale information, into the corresponding decoder. The inputs to the CFA are the $X_0$ and the hierarchical outputs $F_1$, $F_2$ and $F_3$ from the corresponding MWCN blocks.

Given an input hierarchical feature $F_i \in \mathbb{R}^{H\times W \times C}$, the first step is to partition the feature map into patches of sizes $s$ and $2s$, resulting in feature maps $F_i^{s} \in \mathbb{R}^{\frac{H}{s} \times \frac{W}{s} \times s^2 \cdot C}$ and $F_i^{2s} \in \mathbb{R}^{\frac{H}{2s} \times \frac{W}{2s} \times 4s^2 \cdot C}$. In our experiments, $s$ is set to 2. Subsequently, $F_i$, $F_i^{s}$ and $F_i^{2s}$ are all passed through an Average Pooling layer to average the features along the $H$ and $W$ dimensions. For $F_i^{s}$ and $F_i^{2s}$,  the channel features after pooling are randomly grouped and summed according to $s$ and $2s$, integrating local features and reducing dimensionality. Then the feature maps from the three scales are concatenated, shuffled and averaged along the channel dimension. The output is passed through a pointwise convolution layer and a sigmoid function to obtain the attention map $A \in \mathbb{R}^{1 \times 1 \times C}$, as shown in equation \ref{eq2}. 

\begin{equation}
A = \text{Sigmoid}(\text{Conv}_{1\times1}(\text{Mean}(\text{Channel Shuffle}(\text{Concate}(F_i^{'}, F_i^{s'}, F_i^{2s'}))))
\label{eq2}
\end{equation}
Here $F_i^{'}$, $F_i^{s'}$ and $F_i^{2s'}$, represent the output after grouping and summing the three scales channel features. Then, $A$ is multiplied with the original features to complete the channel weighting under multi-scale information, as shown in equation \ref{eq3}, $\hat{F_i}$ represents the output of the CFA block.

\begin{equation}
\hat{F_i} = A \times F_i
\label{eq3}
\end{equation}

\section{Experiments}
\subsection{Implementation Details}
We randomly selected 80\% of the PRAD-10K images as the training set and used the remaining images as the test set. Network parameters were randomly initialized, and the Adam optimizer was used with an initial learning rate of 0.0001 and a 'Poly' learning rate decay strategy. The loss function is a combination of Cross-Entropy loss and DICE loss. The input images were RGB format and was resized to 256×256, the number of training epochs was 200, and the batch size was set to 12. The number of hierarchical feature channels was set to [64, 128, 256, 512], and number $L_i$ of MWCN blocks was set to [1, 1, 2, 1]. All training was conducted on two NVIDIA GeForce RTX 3090 GPUs using the PyTorch library. All experiments were repeated three times, and the average results were reported.

\subsection{Comparisons with Other Methods}
To validate the effectiveness of PRNet, we selected both typical and recently proposed SOTA medical image segmentation models for comparison experiments. Table \ref{tab2} presents the quantitative comparisons of PRNet with other SOTA medical image segmentation methods on the PRAD-10K dataset. Overall, PRNet achieved the best performance, with an average DSC of 84.28\%. For some of the recently proposed medical image segmentation models, such as ACC-Unet (MICCAI'23), AHGNN (MICCAI'24) and EMCAD (CVPR'24), PRNet also demonstrated certain advantages, surpassing them by 8.78\%, 4\% and 5.85\% in overall performance, respectively. For dental crowns and implants, their similar features often lead to confusion. However, PRNet outperformed other networks in distinguishing them, achieving the best and second-best segmentation results. Additionally, PRNet significantly outperformed other methods in the recognition of pulp, orthodontic devices, and apical periodontitis, further demonstrating its superior ability to identify small targets and highlighting its capability to address multi-scale challenges. 

\begin{table}[t]
\centering
\caption{Quantitative comparison results of PRNet and other SOTA medical image segmentation methods on the PRAD-10K dataset, with DSC as the evaluation metric. Red indicates the best performance, and blue indicates the second best. RCF, DC, DF, IM, OD and AP represent Root Canal Filling, Denture Crown, Dental Fillings, Implant, Orthodonic Devices and Apical Periodontitis, respectively.}
\begin{tabularx}{\textwidth}{@{}>{\centering\arraybackslash}p{2.5cm} *{9}{>{\centering\arraybackslash}X}|c@{}}
\hline
Model & Tooth & Bone & Pulp & RCF & DC & DF & IM & OD & AP & \textbf{Avg.} \\ \hline
Unet\cite{ronneberger2015u} & 91.55 & 92.17 & \textcolor{blue}{85.74} & 84.82 & 55.33 & 75.35 & 63.69 & 89.71 & 84.61 & 80.33 \\
Unet++\cite{zhou2019unet++} & 92.56 & 93.20 & 85.52 & \textcolor{blue}{85.31} & 57.94 & \textcolor{blue}{78.07} & \textcolor{red}{64.70} & \textcolor{blue}{90.38} & 72.47 & 80.02 \\
Atten-Unet\cite{oktay2018attention} & \textcolor{blue}{92.58} & \textcolor{blue}{93.31} & 85.61 & 84.54 & \textcolor{blue}{69.14} & 75.19 & 63.63 & 89.14 & 84.25 & \textcolor{blue}{81.93} \\
MultiResUnet\cite{ibtehaz2020multiresunet} & 91.70 & 92.64 & 79.07 & 82.58 & 45.16 & 66.97 & 61.53 & 76.02 & 82.01 & 75.31 \\
TransUnet\cite{chen2021transunet} & 91.34 & 92.05 & 75.23 & \textcolor{red}{87.93} & 81.50 & 59.28 & 61.16 & 90.19 & 74.01 & 79.19 \\
Swin-Unet\cite{cao2022swin} & 89.53 & 90.43 & 77.17 & 69.16 & 42.35 & 63.03 & 56.01 & 75.67 & 72.84 & 70.69 \\
UNEXT\cite{valanarasu2022unext} & 90.41 & 91.66 & 74.22 & 55.71 & 47.64 & 64.12 & 56.89 & 72.09 & 76.09 & 69.87 \\
MGFuseSeg\cite{xu2023mgfuseseg} & 91.58 & 92.50 & 84.89 & 73.27 & 48.33 & 67.96 & 62.42 & 70.01 & 75.54 & 73.94 \\
ACC-Unet\cite{ibtehaz2023acc} & 91.69 & 92.68 & 83.34 & 78.49 & 41.10 & 71.68 & 63.27 & 79.98 & 76.92 & 75.46 \\
EMCAD\cite{rahman2024emcad} & 91.62 & 92.40 & 81.98 & 80.44 & 63.26 & 69.52 & 62.36 & 83.48 & 80.41 & 78.39 \\
TinyUnet\cite{chen2024tinyu} & 88.62 & 89.92 & 63.47 & 71.37 & 40.09 & 43.69 & 55.27 & 68.19 & 88.64 & 67.60 \\
AHGNN\cite{chai2024novel} & \textcolor{red}{92.60} & \textcolor{red}{93.45} & 85.60 & 82.49 & 57.31 & 75.24 & 63.78 & \textcolor{blue}{88.49} & 83.16 & 80.24 \\
\hline
PRNet (ours) & 92.38 & 93.02 & \textcolor{red}{88.87} & 82.89 & \textcolor{red}{78.88} & \textcolor{red}{78.66} & \textcolor{blue}{63.92} & \textcolor{red}{92.46} & \textcolor{red}{88.83} & \textcolor{red}{84.24} \\
\hline
\label{tab2}
\end{tabularx}
\end{table}

\subsection{Ablation experiments}
\begin{table}[h]
\setlength{\tabcolsep}{1.5pt} 
        \centering
        \small
        \caption{The quantitative results of the ablation experiments, $\checkmark$ indicates inclusion and $\circ$ indicates exclusion.}
        \label{tab6}
        \begin{tabular}{|c||cccc||c|}
            \hline
              & CFA & MWCN(k=3) & MWCN(k=5) & GFWM & DSC$\uparrow$ \\
            \hline
            \multirow{2}{*}{UNet} & $\circ$ & $\circ$ & $\circ$& $\circ$ & 80.33 \\
            & $\checkmark$ & $\circ$ & $\circ$& $\circ$ & 82.89 \\
            
            \cline{1-5}
            
            \multirow{3}{*}{UNet Decoder}& $\checkmark$ & $\checkmark$ & $\circ$ & $\circ$ & 83.65 \\
            
            &$\checkmark$ & $\circ$ & $\checkmark$ & $\circ$ & 83.43 \\
            
            & $\checkmark$ & $\checkmark$ & $\checkmark$ & $\circ$ & 83.81 \\
            \cline{1-5}
            PRNet & $\checkmark$ & $\checkmark$ & $\checkmark$ & $\checkmark$ & \textbf{84.23}\\
            \hline
        \end{tabular}
\label{tab3}
\end{table}
We conducted ablation experiments to verify the effectiveness of each component of PRNet with results in Table \ref{tab3}. As seen from the second row, adding CFA blocks to the vanilla UNet's skip connections improved segmentation performance,increasing the average DSC by 2.56\%. Next, we replaced vanilla UNet encoder with our MWCN encoder, keeping the decoder structure and CFA block unchanged. To validate the effectiveness of two-scale MWCN encoder, we conducted three comparative experiments. In the third and fourth rows of Table \ref{tab3}, using MWCN encoders with kernel sizes of 3 and 5, respectively, both improved segmentation performance compared to the vanilla UNet encoder. In the fifth row, using two-scale MWCN provided further improvement over single-scale MWCN. Finally, in the last row of the Table \ref{tab3}, we introduced the GFWM into the MWCN, forming our proposed PRNet. From the segmentation results, it can be seen that PRNet achieved the best average DSC, demonstrating the effectiveness of the GFWM. The ablation experiments prove that each component of PRNet interacts synergistically and is indispensable.

\section{Conclusion}
In this paper, we first introduce PRAD-10K, which to the best of our knowledge is the first and largest high-quality expert-annotated dataset for Intelligent analysis of PR in endodontics. We hope that the introduction of PRAD-10K will effectively advance research in this field. Secondly, to establish a benchmark on PRAD-10K, we propose PRNet, a deep network with MWCN encoders and CFA mechanisms. Experimental results demonstrate that PRNet achieves competitive performance on PRAD-10K. Future research plans involve further expansion and refinement of the PRAD dataset, alongside the design and optimization of more efficient fully supervised, semi-supervised, or multimodal PR analysis models. 

\bibliographystyle{splncs04}
\bibliography{./ref.bib}
\end{document}